\documentclass[prb,aps,amsmath,11pt,a4paper]{iopart}
\usepackage{amsfonts,epsfig}
\usepackage{graphicx}

\begin{document}

\title{Supersymmetric Virial Expansion for Time-Reversal Invariant Disordered Systems}

\author{S. Kronm\"{u}ller, O.M. Yevtushenko}
\address{Ludwig Maximilians University, Arnold Sommerfeld Center and Center for Nano-Science, Munich, D-80333, Germany}
\eads{\mailto{sebastian.kronmueller@physik.uni-muenchen.de}, \\
           \mailto{oleg.yevtushenko@physik.uni-muenchen.de}}

\author{E. Cuevas}
\address{Departamento de F\'{i}sica, Universidad de Murcia, E30071 Murcia, Spain}
\eads{\mailto{ecr@um.es}}

\date{\today}

\begin{abstract}
We develop a supersymmetric virial expansion for two point correlation functions of almost diagonal 
Gaussian Random Matrix Ensembles (ADRMT) of the orthogonal symmetry. These ensembles have 
multiple applications in physics and can be used to study universal properties of time-reversal invariant 
disordered systems which are either insulators or close to the Anderson localization transition. We derive 
a two-level contribution to the correlation functions of the generic ADRMT and apply these results to 
the critical (multifractal) power law banded ADRMT. Analytical results are compared with numerical ones.
\end{abstract}


\maketitle


\section{Introduction}
Random matrix theories (RMT) have proven to be useful mathematical models to study quantum mechanical 
disordered or chaotic physical systems \cite{Guhr-1998}. The Wigner-Dyson RMT (WDRMT), which is characterized 
by independent Gaussian probability distributions of the matrix elements with constant variances of the off-diagonal
matrix entries, can be used to describe the statistics of electrons in small disordered samples in the universal (metallic) 
limit, as the supersymmetric $\sigma$-model obtained for the WDRMT is the same as that of electrons in a metallic dot 
with infinite conductance \cite{Efetov-1983,Efetov-1997}.  The ensembles in the WDRMT can be classified by their 
symmetries, for example, the Gaussian orthogonal ensemble (GOE) consists of real symmetric matrices and can be 
used to describe time-reversal invariant systems; the Gaussian unitary ensemble (GUE) consists of Hermitian matrices 
with equal variances of the real and imaginary parts of the off-diagonal matrix elements and can be used to describe systems with 
broken time-reversal symmetry \cite{Mehta-2004}. 

Beside the universal metallic regime, physical applications require RMT models which can be used in cases of
either insulators or critical systems at (or close to) the Anderson localization transition \cite{evers-2008}.
In such unconventional RMTs, the variance of the matrix elements depends on their distance to the diagonal. 
For example, the power law banded RMT (PLBRMT) is an unconventional RMT where the off-diagonal variances decrease
as a power law $r^{-2\alpha}$ of their distance to the diagonal $r$ outside of a band of width $B$ \cite{Mirlin-1996}.  
For $\alpha=1$, the PLBRMT exhibits multifractal eigenfunctions and critical level statistics, a behaviour 
different to the Poisson statistics of ideal insulators and the Wigner-Dyson statistics of ideal metals 
\cite{evers-2008}. Numerical studies reveal an excellent agreement between different correlation functions
of the critical PLBRMT and of the Anderson model at the transition point \cite{Cuevas-PRB-2007,Nishigaki-1999,Zharekeshev-1997}.
The bandwidth controls the fractality of the RMT eigenstates: their fractal dimensions are close to the space 
dimension, $ \, d $, if the bandwidth is large and are much smaller than $ \, d \, $ in the case of the almost
diagonal critical PLBRMT with small bandwidth. The latter regime of strong multifractality is relevant
for the Anderson model in high dimensions.

The methods which can be used for an analytical study of the PLBRMT depend on the bandwidth. For large bandwidths, 
a field theoretical technique of the supersymmetric $\sigma$-model can be applied \cite{Mirlin-1996}. In the opposite 
case of the almost diagonal PLBRMT with small bandwidth,  a complementary method, the supersymmetric virial 
expansion (VE), can be used. The VE is a regular expansion of the correlation functions in a number of interacting energy 
levels. In the supersymmetric version, it corresponds to the formal expansion in a number of independent supermatrices 
with broken supersymmetry. It is the supersymmetric analogy to the linked cluster expansion used to calculate thermodynamic 
quantities for classical imperfect gases \cite{Wannier-1966}.

The VE has been initially developed on the formal basis of the Trotter formula and a combinatorial analysis for a spectral 
form-factor of the generic case of almost diagonal random matrices (ADRMT) from the GUE symmetry class 
\cite{Yevtushenko-JPA-2003}. It has then been extended to calculate the density-of-states and the level compressibility 
\cite{Yevtushenko-PRE-2004,Kravtsov-JPA-2006}. The supersymmetric (SuSy) VE has been developed later for 
unitary ADRMTs \cite{Yevtushenko-JPA-2007}. It is free of the complicated combinatorics and allows one to 
investigate not only the spectral statistics but also the statistics of the eigenfunctions. Recently, the SuSy VE has been used 
to study the dynamical scaling in multifractal systems \cite{Yevtushenko-2009}. 

To use the SuSy VE in the context of physical systems with time-reversal invariance it is necessary to modify it such that it is applicable to orthogonal ADRMTs. 
Many experimentally relevant disordered quantum systems, e.g. cold quantum gases in a disordered optical lattice, belong to this class of systems  \cite{Billy-2008,Roati-2008}.
This paper fills this gap in the theory of the ADRMTs, namely, we extend the general formalism of the supersymmetric VE to orthogonal ADRMTs. 

We organise the paper as follows: basic definitions of the almost diagonal RMT and of the two point correlation functions 
are given in Section 2; the supersymmetric representation of the correlation functions and of the supersymmetric VE are explained in 
Sections 3 and 4 respectively; a parametrization of supermatrices is given in Section 5; generic results for the leading terms of the 
VE are calculated in Section 6 and are applied to study the local density-of-states (LDOS) of the almost diagonal critical PLBRMT in 
Section 7. To the best of our knowledge, the LDOS-LDOS correlation function of the critical orthogonal PLBRMT was 
never calculated before. We compare the analytical results with the numerical ones and conclude the paper with a
brief discussion.

\section{Main Definitions}
We consider a Gaussian real symmetric ADRMT defined by the following statistics of matrix entries:
\begin{equation}
	\langle H_{ij}\rangle=0,	
		\quad 
			\langle H_{ii}^{2}\rangle=1,	
				\quad 
					\langle H_{i\neq j}^{2}\rangle=B^{2}F(|i-j|)\equiv b_{ij} \ll 1 \, ,
\label{RMTVar_Val}
\end{equation}
where $F(x)$ is a real decaying function and the small parameter $B\ll1$ reflects that the ensemble is almost diagonal. 
The matrices of the ensemble are of size $N\times N$ with $N\gg 1$. We introduce the retarded and advanced Green's 
functions by
\begin{equation}
	\hat{G}^{R/A}(E)
		\equiv
			\frac{1}{E-\hat{H}\pm\mathit{i}\eta}, \ \eta \to +0 \, ;
\label{Green_Def}
\end{equation}
with $\hat{H}$ being an arbitrary matrix of the ensemble and we define the two point Green's function at the band 
center $E=0$ as
\begin{equation}
	\mathcal{G}_{pq}(\omega)
		\equiv 
			\hat{G}^{R}_{pp}\left(\frac{\omega}{2}\right)\hat{G}^{A}_{qq}\left(-\frac{\omega}{2}\right), 
				\label{TPGreen_Def}
\end{equation}
using the matrix elements of the Green's functions $\hat{G}^{R/A}_{nn}=\langle n|\hat{G}^{R/A}|n\rangle$ ( $|n\rangle$ with $n=1\ldots N$ is a canonical basis vector in the vector space of the RMT ).

We will consider disorder averaged two point-correlation functions, namely, the spectral correlation function and the diagonal 
local-density-of-states (LDOS) correlation function, which are defined by
\begin{equation} 
	R_2(\omega) 
		\equiv 
			\Delta^2 \sum_{i,j=1}^N \left\langle\left\langle
									\delta\left(E_i+\frac{\omega}{2}\right) \delta\left(E_j-\frac{\omega}{2}\right) 
							     \right\rangle\right\rangle, 
			\label{SpecCor_Def}
\end{equation}
and
\begin{equation}
	C_2(\omega) 
		\equiv 
			\Delta^2\sum_{p=1}^N \sum_{i,j=1}^N 
										\left\langle\left\langle
											 \left|\psi_i(p)\right|^2 \left|\psi_j(p)\right|^2 \delta\left(E_i+\frac{\omega}{2}\right) \delta\left(E_j-\frac{\omega}{2}\right) 
								   		\right\rangle\right\rangle.
			\label{LDOSCor_Def}
\end{equation}
Here $\Delta$ is the mean level spacing of the RMT; $E_{i,j}$ are the eigenvalues of the random matrices; $ \, \langle\langle 
ab \rangle\rangle \equiv \langle ab\rangle-\langle a\rangle\langle b\rangle \, $ and $\langle\ldots\rangle$ denotes the average over 
the ensemble of random matrices. For orthogonal almost diagonal RMT, the mean level spacing is known to be 
\cite{Yevtushenko-JPA-2003}:
\begin{equation}
	\Delta\big|_{B\ll1}
		\approx 
			\frac{1}{N}\left(\sqrt{2\pi}+O\left(B^2\right)\right). 
				\label{Mls_Val}
\end{equation}
The spectral and the LDOS correlation functions can be expressed by the two point Green's function:
\begin{eqnarray} 
   R_2(\omega)  = \frac{\Delta^2}{2\pi^2} \sum_{p,q=1}^N 
			\mathrm{Re}\left[\langle\langle\mathcal{G}_{pq}(\omega)\rangle\rangle\right], 
\label{SpecCor_Green} \\
   C_2(\omega) = \frac{\Delta^2}{2\pi^2}\sum_{p=1}^N 
			\mathrm{Re}\left[\langle\langle\mathcal{G}_{pp}(\omega)\rangle\rangle\right].
\label{LDOSCor_Green}
\end{eqnarray}
Thus, one needs $ \, \langle\langle\mathcal{G}_{pq}(\omega)\rangle\rangle \, $ to study the correlation functions 
(\ref{SpecCor_Def},\ref{LDOSCor_Def}).


\section{Supersymmetric Representation of the Two Point Green's Function}

The ensemble averaged two point Green's function can be obtained with the help of a supersymmetric field theory \cite{Yevtushenko-JPA-2007}. Let us introduce $2N$ super vectors of the form
\begin{eqnarray}
	\psi_i^{R/A}
			\equiv
				\left(\begin{array}{c} S_i^{R/A} \\ \chi_i^{R/A} \\ \left(S_i^{R/A}\right)^* \\ \left(\chi_i^{R/A}\right)^*\end{array}\right), \\
	\left(\psi_i^{R/A}\right)^{\dagger}
							\equiv
								\left( \left(S_i^{R/A}\right)^*, \left(\chi_i^{R/A}\right)^*,S_i^{R/A}, -\chi_i^{R/A} \right),\label{svectors}
\end{eqnarray}
where $i=1,\ldots,N$, $S_i^{R/A}$ are commuting and $\chi_i^{R/A}$ are anti-commuting Grassmannian variables and the indices $R/A$ stand for the retarded/advanced sectors. We use the outer product of these super vectors to define the supermatrices
\begin{equation}
	Q_i
		\equiv
			\frac{1}{2}\left(\begin{array}{cc} 
									\psi_i^{R}\otimes\left(\psi_i^R\right)^{\dagger} & - \psi_i^{R}\otimes\left(\psi_i^A\right)^{\dagger} \\
									\psi_i^{A}\otimes\left(\psi_i^R\right)^{\dagger} &  - \psi_i^{A}\otimes\left(\psi_i^A\right)^{\dagger} 
			\end{array}\right).\label{SupMatrixDef}
\end{equation}
These definitions allow one to write the ensemble averaged two point Green's function as
\begin{equation}
	\left\langle \mathcal{G}_{pq}(\omega) \right\rangle 
									= 
									\int D \{Q\} \mathcal{R}_p \mathcal{A}_q 
										\left(\prod_{i=1}^N e^{S_0[Q_i]}\right)\left(\prod_{i\neq j}^N e^{S_p[Q_i,Q_j]}\right). 
											\label{Green_Int}
\end{equation}
Here $\int D \{Q\}\equiv\prod_{i=1}^N D\{Q_i\}$ is the measure of integration over the supermatrices $Q_i$ and $\mathcal{R}_p$, 
$\mathcal{A}_q$ are the symmetry-breaking factors between the commuting and anti-commuting variables in the retarded and advanced 
sectors, respectively. They can be chosen as follows
\begin{equation}
	\mathcal{R}_p 
				= \chi_p^R \left( \chi_p^R\right)^*,
		\quad
	\mathcal{A}_q
				=  \chi_q^A \left( \chi_q^A\right)^*.
\end{equation}
The action of the field integral is split in two parts: $S_0$ depends on a single supermatrix
\begin{equation}
	S_0[Q_i]
			\equiv 
				- \frac{1}{2} \left(\mathrm{Str}[Q_i]\right)^2+i\frac{\omega+i\eta}{2}\mathrm{Str}[\Lambda Q_i],
				\quad 
					\Lambda 
						\equiv
							\left(\begin{array}{cc} 1 & 0 \\ 0 & -1 \end{array}\right),
\end{equation}
$S_p$ depends on the product of two supermatrices and is proportional to the off-diagonal variances
\begin{equation}
	S_p[Q_i,Q_j]
			\equiv 
				- b_{ij} \mathrm{Str}[Q_iQ_j]. 
					\label{Two_Matrix_Action}
\end{equation}


\section{The Virial Expansion}
The two point Green's function \eref{Green_Int} of the ADRMT can be approximately calculated using
the virial expansion \cite{Yevtushenko-JPA-2003,Yevtushenko-PRE-2004,Kravtsov-JPA-2006,Yevtushenko-JPA-2007}. 
The basic concept of the VE is similar to the idea of resonant level interactions which has been used in a semi-empirical 
renormalization group approach \cite{Levitov-PRL-1990,Levitov-ADP-1999}: a small probability for energy levels of
the ADRMT to interact in the energy space allows one to develop a regular perturbative expansion in the number of
interacting levels. Each term in the VE reflects a contribution of a certain number of simultaneously interacting
levels. In contrast to the semi-empirical renormalization group approach, the VE is a regular perturbative expansion, 
thus allowing to include an arbitrary number of resonant level interactions and a controlled estimate of the contribution
of the unused resonances.

The mathematical expressions of the virial expansion can be obtained by rewriting the term containing the off-diagonal 
variances in the superintegral \eref{Green_Int}
\begin{equation}\label{FormalVE}
	\prod_{i\neq j}^N e^{S_p[Q_i,Q_j]}
				\equiv \mathcal{V}^D + \sum_{m=2}^{\infty} \mathcal{V}^{(m)} \, ,
\end{equation}
\begin{equation}\label{M-functions}
\mathcal{V}^D = 1 \, ,
	\quad 
\mathcal{V}^{(2)} = \sum_{i>j=1}^N \left( e^{2S_p[Q_i,Q_j]}-1 \right) \, ;
\end{equation}
and so on (see details in \cite{Yevtushenko-JPA-2007}). 

The VE is similar to the cluster expansion in the classical gas theory \cite{Wannier-1966}, however, the SuSy representation of
the resulting terms ensures that, in contrast to the cluster expansion in the classical gas theory, no further reordering of the 
terms is necessary. Thus, $ \, {\cal V}^{D} \, $ corresponds to the contributions of the diagonal part of the RMT,  $ \, 
{\cal V}^{(2)} \, $ reflects the contributions of two-level interactions and so on. Using Eqs. (\ref{FormalVE}-\ref{M-functions}), 
we can formally expand the averaged two point Green's function:
\[
    \langle {\cal G}_{pq} \rangle = \left\langle \mathcal{G}_{pq}^D \right\rangle + 
				\left\langle \mathcal{G}_{pq}^{(2)} \right\rangle + \ldots
\]
After integrating out all supermatrices $ \, Q_j \, $ which neither included in the SuSy breaking factor ($ j \ne p, q $)
nor are linked to this factor through $ \, {\cal V}^{(m)} \, $ functions \cite{Yevtushenko-JPA-2007}, the first two terms 
of the VE read
\begin{equation}
	 \left\langle \mathcal{G}_{pq}^D \right\rangle 
	 	=  \int D\{Q_p\} D\{Q_q\} \mathcal{R}_p \mathcal{A}_q e^{S_0[Q_p]+S_0[Q_q]},
\label{diag_supint}
\end{equation}
\begin{eqnarray}
\fl	\left\langle \mathcal{G}_{p\neq q}^{(2)} \right\rangle 
		=  \int D\{Q_p\} D\{Q_q\} \mathcal{R}_p \mathcal{A}_q 
			\left( e^{S_0[Q_p]+S_0[Q_q]} \right) \left( e^{2S_p[Q_p,Q_q]} - 1 \right),
\label{sec_supint_nondiag} \\
\fl	\left\langle \mathcal{G}_{pp}^{(2)} \right\rangle 
			=  \sum_{n\neq p}^N \int D\{Q_p\} D\{Q_n\} \mathcal{R}_p \mathcal{A}_p
				\left( e^{S_0[Q_p]+S_0[Q_n]} \right) \left( e^{2S_p[Q_p,Q_n]} - 1 \right) .
\label{sec_supint_diag}
\end{eqnarray}


\section{Parametrization of the Supermatrices}
A convenient parametrization of the $ \, Q $-matrices for the SuSy VE has been first suggested in \cite{Ossipov-PRB-2006}.
For each supermatrix, one introduces 2 positive variables $ \, \lambda_i^{R/A} \in [0,\infty] $, 2 angles $ \, \phi_i^{R/A} \in 
[0,2\pi] \, $ and 4 anti-commuting Grassmann variables $\eta_i^{R/A}$, $\left(\eta_i^{R/A}\right)^*$ ($i=1,\ldots N$)
in the following way:
\begin{equation}
\fl	S_i^{R/A}
		= \lambda_i^{R/A} \left( 1 - \frac{1}{2} \left( \eta_i^{R/A} \right)^* \eta_i^{R/A} \right) e^{i\phi_i^{R/A}}, 
	\quad
	\chi_i^{R/A} 
		= \lambda_i^{R/A} \eta_i^{R/A} e^{i\phi_i^{R/A}}. 
\label{parametrization}
\end{equation}
The integration measure in this parametrization is given by
\begin{equation}
     D\{Q\}
	= \prod_{i=1}^N \frac{ d\lambda_i^R d\phi_i^R \left( d\eta_i^{R} \right)^* d\eta_i^{R} }{ \pi \lambda_i^R } 
			\frac{ d\lambda_i^A d \phi_i^A  \left( d\eta_i^{A} \right)^* d\eta_i^{A} }{ \pi \lambda_i^A },
\end{equation}
and the single matrix part of the action reads
\begin{equation}
	S_0[Q_i]
	= - \frac{1}{2} \left( \left( \lambda_i^R \right)^2 - \left( \lambda_i^A \right)^2 \right)^2
		+ i \frac{ \omega + i \eta }{ 2 } \left( \left( \lambda_i^R \right)^2 + \left( \lambda_i^A \right)^2 \right).
\label{single_action}
\end{equation}
Using rotations in superspace, we can find a convenient representation of the two matrix part of the action in the new variables 
(see \ref{ActionSimplification} for details).
\begin{eqnarray}
	& S_p[Q_i,Q_j] = 
		& -b_{ij} \left( \lambda_i^R \lambda_j^R \cos \left( \theta_{ij}^R \right) 
			\left( 1 - \frac{1}{2} \left( \alpha_{ij}^R \right)^* \alpha_{ij}^R \right) - \right. \nonumber \\ 
	& 			   
		& \left. - \lambda_i^A \lambda_j^A \cos \left( \theta_{ij}^A \right) 
			\left( 1 - \frac{1}{2} \left( \alpha_{ij}^A \right)^* \alpha_{ij}^A \right) \right)^2. 
\label{Simpl_Two_Matrix_action}
\end{eqnarray}
where $\alpha_{ij}^{R/A}\equiv \eta_i^{R/A}-\eta_j^{R/A}$ and
\begin{equation}
	\theta_{ij}^{R/A} \equiv \phi_j^{R/A} - \phi_i^{R/A} - \frac{i}{2} 
		\left( \left( \eta_i^{R/A} \right)^* \eta_j^{R/A} - \left( \eta_j^{R/A} \right)^* \eta_i^{R/A} \right).
\end{equation}
As a direct consequence of the underlying symmetries of the ensembles, the supermatrices defined in Eq. \eref{SupMatrixDef} are twice the size of the supermatrices which have been used for the SuSy VE of the unitary ADRMT ( cf. Eq. (A.1),(A.2) \cite{Yevtushenko-JPA-2007} ). Parametrizing the supermatrices as defined in Eq. \eref{parametrization}, this yields a different amount of effective variables necessary for the integral representation of the two point Green's function; the two matrix part of the action contains one effective angle in the case of the GUE ( cf. Eq. (A.10) \cite{Yevtushenko-JPA-2007}), whereas for the GOE two effective angles ($\theta_{ij}^R$ and $\theta_{ij}^A$ in \eref{Simpl_Two_Matrix_action}) appear.
For the further calculations, we will use other commuting variables which allow one to simplify the single matrix part of the action:
\begin{equation}
	R_i
		\equiv
			\left( \lambda_i^R \right)^2 - \left( \lambda_i^A \right)^2,
				\quad 
					S_i
						\equiv
							\left( \lambda_i^R \right)^2 + \left( \lambda_i^A \right)^2 \, ;
\end{equation}
where $ \, -S_i \le R_i \le S_i $, $ \, S_i \in [0,\infty] $. The integration measure changes to
\begin{equation}
	D\{Q\}
		=
		\prod_{i=1}^N \frac{ dS_i dR_i d\phi_i^R d\phi_i^A \left( d\eta_i^{R} \right)^* d\eta_i^{R} \left( d\eta_i^{A} \right)^* d\eta_i^{A} }{ 2 \pi^2 \left( S_i^2 - R_i^2 \right) }.
\end{equation}


\section{Generic Results for the First Two Terms of the Virial Expansion}
Integration of the superintegral in \eref{diag_supint} and calculation of the diagonal contribution of the product of single averaged Green's functions $\left<\hat{G}^R\left(\omega/2\right)\right>\left<\hat{G}^A\left(-\omega/2\right)\right>$ yields:
\begin{equation}
	\left\langle \left\langle \mathcal{G}_{p\neq q}^D \left( \omega\right ) \right \rangle \right \rangle = 0, 
		\quad
	\left\langle \left\langle \mathcal{G}_{pp}^D \left( \omega \right) \right\rangle \right\rangle \Big|_{N\gg1} 
		\approx \frac{2Ni}{s+\frac{i\eta}{\Delta}}=2\pi N \delta(s)+\frac{2Ni}{s}, 
\end{equation}
where $s=\omega/\Delta$. 

The superintegrals \eref{sec_supint_nondiag}, \eref{sec_supint_diag} can be calculated using a saddle-point approach 
(cf. the Section 4.1 in \cite{Yevtushenko-JPA-2007}). 
Rescaling all $S_i$ and $R_i$ variables with $\bar{S}_i=\sqrt{b_{p\alpha}}S_i$, $\bar{R}_i=\sqrt{b_{p\alpha}}R_i$, where $\alpha\in\{n,q\}$, 
we obtain a large negative factor of order $ \, O(1/B^2) \, $ in front of $ \, R_i^2 \, $ in the single matrix part of the action
\begin{equation}
	S_0[Q_i] = - \frac{1}{2 b_{p\alpha} } \bar{R}_i^2 + i \frac{ \omega + i \eta }{ 2 \sqrt{b_{p\alpha}} } \bar{S}_i \, , \quad
			b_{p \alpha} \ll 1 \, .
\label{single_action_RS}
\end{equation}
This allows to employ a saddle point approach in the $R_i$:
\begin{eqnarray}
\fl	\int_0^{\infty} d\bar{S}_{ p , \alpha } \int_{ - \bar{S}_{p,\alpha} }^{ \bar{S}_{ p , \alpha} } d\bar{R}_{ p , \alpha}
		\frac{ \mathcal{R}_p \mathcal{A}_{\alpha} 
			f \left( \bar{R}_{ p , \alpha } , \bar{S}_{ p , \alpha } \right) }{ 
					\left( \bar{S}_p^2 - \bar{R}_p^2 \right) 
						\left( \bar{S}_\alpha^2 - \bar{R}_\alpha^2 \right) }
			\left(e^{-\frac{1}{2 b_{p\alpha}} \left( \bar{R}_p^2+ \bar{R}_\alpha^2 \right) + i \frac{ \omega + i \eta}{ 2 \sqrt{b_{p\alpha}} } \left( \bar{S}_p + \bar{S}_\alpha\right) } \right) \nonumber \\
\fl \approx 
	2 \pi b_{p\alpha} \Bigg( \int_0^{\infty} d\bar{S}_{p,\alpha}
	\frac{ \mathcal{R}_p \mathcal{A}_{\alpha}
			f \left( 0 , \bar{S}_{p,\alpha} \right)}{ \bar{S}_p^2 \bar{S}_\alpha^2}
				\left( e^{i\frac{\omega + i\eta}{ 2 \sqrt{b_{p\alpha}} } \left( \bar{S}_p + \bar{S}_\alpha \right)} \right) \nonumber
					+ O \left( \sqrt{b_{p\alpha}} \right) + O \left( \frac{ \bar{R}_t }{ \bar{S}_t} \right) \Bigg),
									\label{saddle_point}
\end{eqnarray}
where $f\left(\bar{R}_{p,\alpha},\bar{S}_{p,\alpha}\right)\equiv \left(e^{2S_p[Q_p,Q_n]}-1\right)$ and $\bar{R}_t$, $\bar{S}_t$ are the typical 
values where the integrals over $\bar{R}$ and $\bar{S}$ converge. Based on rough estimates, one obtains two requirements for the validity 
of the saddle point approximation \cite{Yevtushenko-JPA-2007}:
\begin{equation}
	\omega \le B \ll 1\, .
\end{equation}
However,  the analytical results derived from the saddle point approximation and the results of the  numerical diagonalization are
in a very good agreement even in the regime $ \, \omega \simeq 1 $; see an example of the LDOS-correlation function $ \, 
C(\omega) $, Eq. \eref{C_W_Def}:  \fref{LDOS_fig} below, and figure 3 in the paper \cite{Cuevas-PRB-2007} in the cases of the 
orthogonal and the unitary critical ADRMT, respectively. Corrections beyond the saddle point approximation will be studied in details 
elsewhere \cite{Yevtushenko-2009b}.

After using the saddle point approximation, we can integrate out the angles, the Grassmann variables, and the remaining real variables. This yields
\begin{eqnarray}
\fl	\left\langle \left\langle \mathcal{G}_{p\neq q}^{(2)} \left( \omega \right) \right\rangle \right\rangle 
			\approx \sum_{k=1}^\infty
				 \left( - 1\right)^k 
				 	\left( \frac{ 2 \sqrt{2b_{pq}} }{ i ( \omega + i \eta) } \right)^{2k}
						\frac{ \Gamma^2 \left( k - \frac{1}{2} \right) ( 2k-1 )( k-1 ) }{ 4 \Gamma \left( k + 1 \right) }, 
								\label{nd_sum}\\
\fl	\left\langle \left\langle \mathcal{G}_{pp}^{(2) }\left( \omega \right) \right\rangle \right\rangle 
			\approx \sum_{n\neq p}^N
				\sum_{k=1}^\infty 
					\left( - 1 \right)^k
						\left( \frac{ 2 \sqrt{2b_{pn}} }{ i ( \omega + i \eta ) } \right)^{2k}
							\frac{ \Gamma^2 \left( k - \frac{1}{2} \right) ( 2k - 1 ) }{ 4 \Gamma \left( k \right) }.
			 					\label{d_sum}
\end{eqnarray}
These power series can be summed up by using a Fourier transform of the physically relevant real part of the two point Green's function: 
\begin{equation}
	\mathrm{Re}\left[\mathcal{G}_{pq}(t)\right]
								=
								\frac{1}{2\Delta}\int_{-\infty}^{\infty}d\omega e^{-i\omega t}
									\left( \mathcal{G}_{pq}\left(\omega \right) + \mathcal{G}_{pq}^*\left(\omega \right)\right). 
										\label{FT}
\end{equation}
After applying the Fourier transform to Eqs. \eref{nd_sum}, \eref{d_sum}, we find the two point Green's function in the time domain:
\begin{eqnarray}
	\mathrm{Re} \left[ \left\langle \left\langle \mathcal{G}_{p\neq q}^{(2)} ( t ) \right\rangle \right\rangle \right] 
							= 
							\frac{ 2 \pi^2 }{ \Delta } b_{pq} |t| e^{-b_{pq} t^2 - \eta|t| }
								I_1 \left( b_{pq} t^2 \right),
													\label{TPG_Ti_ND}\\
	\mathrm{Re} \left[ \left\langle \left\langle \mathcal{G}_{pp}^{(2)} ( t ) \right\rangle \right\rangle \right] 
							= 
							 - \frac{ 2 \pi^2 }{ \Delta } \sum_{n\neq p}^N b_{pn} |t| e^{ - b_{pn} t^2 - \eta|t|}
							 	I_0 \left( b_{pn} t^2 \right).
													\label{TPG_Ti_D}
\end{eqnarray}
The energy representation of the two point Green's function can be obtained by the inverse Fourier transform. If $ \, \omega \ne 0 \, $ we find:
\begin{eqnarray}
	\mathrm{Re} \left[ \left\langle \left\langle \mathcal{G}_{p\neq q}^{(2)} ( \omega ) \right\rangle \right\rangle \right] 
					= \left( \frac{ \pi }{ 2 } \right)^{ \frac{3}{2} } |\bar{\omega}| 
						e^{ - \bar{ \omega }^2 }
							\left( 3 I_0 \left( \bar{ \omega }^2 \right) + I_1 \left( \bar{ \omega }^2 \right) \right)
								- \pi, 
									\label{TPG_En_ND} \\
	\mathrm{Re} \left[ \left\langle \left\langle \mathcal{G}_{pp}^{(2)} ( \omega ) \right\rangle \right\rangle \right] 
					= \sum_{n\neq p}^N
						\left( \frac{ \pi }{ 2 } \right)^{ \frac{3}{2} } |\bar{\omega}|  
							e^{ - \bar{\omega}^2}
								\left( I_0 \left( \bar{ \omega }^2 \right) - I_1 \left( \bar{ \omega }^2 \right) \right);
									\label{TPG_En_D}
\end{eqnarray}
where $\bar{\omega}=\omega/(4\sqrt{b_{p\alpha}})$, and $\alpha=q,n$ in Eqs. \eref{TPG_En_ND} and \eref{TPG_En_D}, respectively. These are the 
generic results for the two matrix approximation of the two point Green's function for ADRMTs of the GOE symmetry class.


\section{Application to the Critical PLBRMT}
The critical PLBRMT is defined as follows:
\begin{equation}
	\langle H_{ij} \rangle=0,	
		\quad 
			\langle H_{ii}^{2} \rangle=1,
				\quad 
					\langle H_{i\neq j}^{2} \rangle = \frac{1}{2} \frac{B^{2}}{B^2+|i-j|^2} \Bigl|_{B \ll 1}
						\approx \frac{1}{2} \frac{B^2}{|i-j|^2} .
\label{CPLBRMTVar_Val}
\end{equation}
The ensemble (\ref{CPLBRMTVar_Val}) is named ``critical'' because it shows multifractal behaviour of the eigenstates and critical level 
statistics at arbitrary $ \, B \, $ \cite{Mirlin-1996}. 

\subsection{The Spectral and the LDOS Correlation Functions}
Inserting the variances \eref{CPLBRMTVar_Val} into Eqs. \eref{SpecCor_Green}, \eref{LDOSCor_Green}, \eref{TPG_En_ND}, 
\eref{TPG_En_D}, we obtain the spectral and the LDOS correlation functions for the critical ADRMT of the orthogonal symmetry: 
\begin{eqnarray} 
	\fl R_2(\omega) 
		=  \frac{1}{N^2} \sum_{n\neq p=1}^N 
		   	\left( \sqrt{2\pi} |\tilde{\omega}| |p-n| e^{-\tilde{\omega}^2|p-n|^2} I_0\left( \tilde{\omega}^2 |p-n|^2 \right)-1\right), 
\label{SpecCor_CPLBRMT}\\
	\fl C_2(\omega) 
		=  \frac{1}{2N^2} \sum_{n\neq p=1}^N
			\sqrt{\frac{\pi}{2}}|\tilde{\omega}||p-n| e^{-\tilde{\omega}^2|p-n|^2}
			\left(I_0\left(\tilde{\omega}^2|p-n|^2\right)-I_1\left(\tilde{\omega}^2|p-n|^2\right)\right),
\label{LDOSCor_CPLBRMT}
\end{eqnarray}
here $\tilde{\omega}=\omega/(\sqrt{8}B)$.
Let us introduce a shifted (positive) spectral correlator:
\begin{equation}
	R_2^+(\omega) \equiv 
		\frac{1}{N^2} \sum_{n\neq p=1}^N  \sqrt{2\pi} |\tilde{\omega}| |p-n| e^{-\tilde{\omega}^2|p-n|^2} I_0
									\left( \tilde{\omega}^2 |p-n|^2 \right) \, ,
\end{equation}
and scale the LDOS correlation function:
\begin{equation}
	C(\omega)
			\equiv
				\frac{C_2(\omega)}{R_2^+(\omega)}. \label{C_W_Def}
\end{equation}
This scaling eliminates the level repulsion effect at $ \, \omega < B \Delta $, therefore, $ \, C(\omega) \, $ is convenient to study the 
correlations of the eigenfunctions taken at the same space point and with a given energy difference. The unitary case has been 
explored in \cite{Cuevas-PRB-2007}. The qualitative behaviour of $ \, C(\omega) \, $ for the orthogonal case is similar to the
unitary one, see \fref{LDOS_fig}. There are three regions which reflect different physical phenomena: 

\begin{figure}
\includegraphics[width=\textwidth]{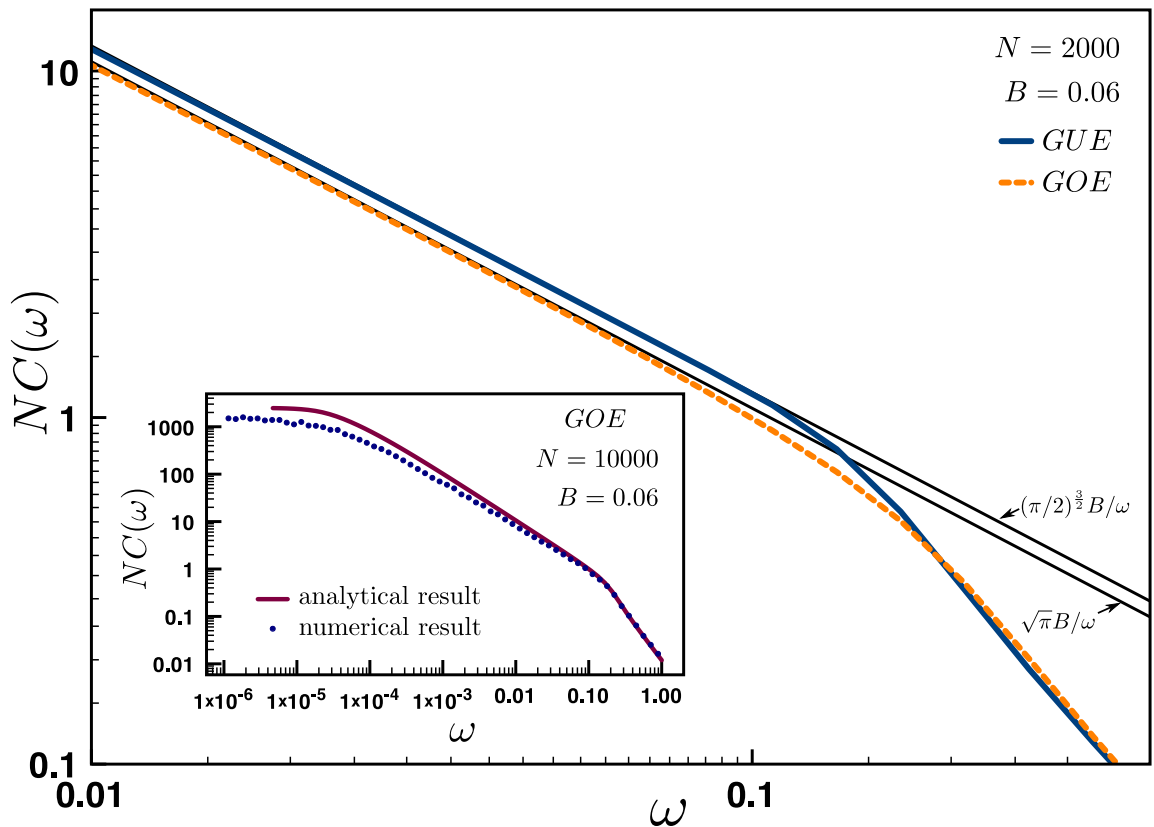}
\caption{\label{LDOS_fig}
(coloured online) 
LDOS correlation function for the critical ADRMT with $ \, B = 0.06 $. 
{\bf Main panel:} Comparison of the analytical results for the cases of GOE and GUE , $ \, N = 2000 $. Black solid lines indicate a 
reference slope predicted by the two matrix approximation of the VE, Eq.(\ref{Scaling})
{\bf Inset:} Comparison of the analytical and numerical results for the case of GOE, $ \, N = 10000 $.  Note that the difference between 
the numerically and the analytically found slope in the intermediate regime $ B \Delta \ll \omega\ll E_0 $ is due to the accuracy of the two 
matrix approximation, see detailed discussion in the main text.
}
\end{figure}

1) $ \, \omega > E_0 \propto B $: the energy difference is larger than the hopping band width of the ADRMT which results in anticorrelations 
of the eigenstates and in a generic power law behaviour $ \, C(\omega) \propto \omega^{-2} \, $ (see \cite{Cuevas-PRB-2007} for details).

2) $ \, 0 < \omega < B \Delta $: the energy separation is very small so that the behaviour of $ \, C_2 \, $ and $ \, R_2 \, $ is governed by
the level repulsion in the energy space, therefore, the ratio $ \, C =  C_2 / R_2 \, $ saturates to a constant. Note the analytically
obtained constant is a bit larger than the numerically found value because of the reasons explained below.  

3) $ \, B \Delta < \omega < E_0 $: this is the most interesting region where strong spatial correlations of the fractal eigenstates must result in 
a scaling $ \, C(\omega) \propto \omega^\mu \, , \ \mu = {d_2/d  - 1} $, $ \, d_2 \, $ and $ \, d \, $ are the second fractal dimension and the
space dimension, respectively \cite{Chalker-1990,Chalker-PRL-1988}.  
This property is especially nontrivial in the case of strong multifractality $ \, d_2 \ll d $ ($ d = 1 \, $ for RMT) where the fractal eigenstates 
are sparse, nevertheless, they are strongly correlated in space. The fractal 
dimension $ \, d_2 \, $ depends on the symmetry class and on the parameter $ \, B $, for example: $ \, d_2 = \sqrt{2}B\, $ for the GOE and  $ \, d_2 = \pi B/\sqrt{2} \,$ for the GUE critical ADRMT ( cf. \cite{Mirlin-2000b} ). Thus,
the critical ADRMT corresponds to the case of the strong multifractality. The parameter $ \, E_0 \, $ can be referred to as ``the upper cut-off of the multifractality''. 
It can be shown from Eq.(\ref{LDOSCor_CPLBRMT}) that
\begin{equation}\label{E0}
   E_0 \simeq \frac{1}{2} N \Delta d_2 \, .
\end{equation} 
Note that the same result holds true for the GUE ensemble: substituting the unitary values of $ \, \Delta \, $ and $ \, d_2 \, $ into Eq.(\ref{E0}) 
we find the unitary value of $ \, E_0 $ \cite{Cuevas-PRB-2007}:
 \begin{equation}
	E_0\Big|_{GUE}=\left(\frac{\pi}{2}\right)^{\frac{3}{2}}B>E_0\Big|_{GOE}=\sqrt{\pi}B.
\end{equation}
If $ \, d_2 \, $ is small, the leading term of the VE cannot distinguish 
the exponents $ \, \mu = d_2  - 1 \, $ and $ \, \mu \simeq -1 $. One has to take into account the interaction of a larger number of energy levels 
to find the correct exponent \cite{Yevtushenko-2009}. That is why the analytical result (\ref{LDOSCor_CPLBRMT}) shows the scaling 
\begin{equation}\label{Scaling}
   NC( B \Delta < \omega < E_0 ) \simeq \frac{E_0}{\omega} 
\end{equation}
see \fref{LDOS_fig}, and the analytically calculated constant for $ \, C(\omega < B \Delta) \, $ is a bit larger as compared with the numerically
obtained value. As $E_0\Big|_{GUE}>E_0\Big|_{GOE}$, the GUE curve in \fref{LDOS_fig} is shifted upwards compared to the GOE curve.


\section{Conclusions}

In this paper we have developed the supersymmetric virial expansion for the two point correlation functions of the almost 
diagonal Gaussian Random Matrix ensembles of the orthogonal symmetry class. We have derived the generic results for 
the two matrix approximation of the two point Green's function in the time (Eqs. \eref{TPG_Ti_ND},  \eref{TPG_Ti_D}) and 
energy (Eqs.  \eref{TPG_En_ND},  \eref{TPG_En_D}) representations. This contribution to the Green's function results from
an interaction of two energy levels in the energy space and can be used to study statistical properties of time-reversal invariant
disordered systems which are either insulators or close to the point of the Anderson localization transition. 

To demonstrate how the method works, we have applied the generic results of the two matrix approximation to the 
Gaussian orthogonal critical power law banded RMT with small bandwidth, Eq. \eref{CPLBRMTVar_Val}. We have obtained for this 
ensemble the spectral correlation function (Eq.  \eref{SpecCor_CPLBRMT}) and the local-density-of-states correlation function 
(Eqs. \eref{LDOSCor_CPLBRMT},\eref{C_W_Def}), the latter has to the best of our knowledge not been derived before. A
comparison of the analytically obtained function $ \, C(\omega) \, $ with the results of the direct numerical diagonalization
shows an excellent agreement at $ \, \omega \ge B $, see the inset in \fref{LDOS_fig}. In the region $ \, B \Delta < \omega < B \, $ the two
matrix approximation yields qualitatively correct results: we find a scaling $ \, C(\omega) \propto \omega^{- 1} \, $ in this
parametrically large energy window, however, the exponent is different from the exponent of the expected critical scaling  
$ \, C(\omega) \propto \omega^{d_2 - 1} \, $ \cite{Chalker-1990}.  This is because the second fractal dimension
is small in the case of the ADRMT, $ \, d_2 \sim B \ll 1 \, $ \cite{Mirlin-2000b} and the leading term of the VE is unable to
distinguish exponents $ \, 1 - d_2 \, $ and $ \, 1 $ \cite{Cuevas-PRB-2007}. One has to take into account the interaction
of a larger number of the energy levels to find the correct exponent \cite{Yevtushenko-2009}. Nevertheless, a
coefficient in the scaling is proportional to $ \, d_2: \quad  C(\omega) \simeq \Delta d_2 / (2\omega) $, see Eq.(\ref{E0},\ref{Scaling}).
Thus, even the results of the two matrix approximation reflect multifractality of the critical eigenstates \cite{Yevtushenko-2009}.
A Fourier transform of $ \, C(\omega)  \, $ yields the return probability for an initially localized wave packet. Therefore, the results
of Section 7 might be relevant to study fractal properties of cold atoms in a disordered optical potential which does
not break the time reversal symmetry \cite{Billy-2008,Roati-2008}.

A generalization of the current results for the three matrix approximation is straightforward (cf. \cite{Yevtushenko-JPA-2007}). In the
future, we plan to use a SuSy field theory, which is the starting point of the of the perturbative VE (Eq. \eref{Green_Int}), to obtain 
nonperturbative results accounting for the interaction of all energy levels at least with an accuracy of the first loop
of the Renormalization Group procedure.

\ack
We would like to thank Vladimir Kravtsov for initiating the project on the field theory representation of the almost diagonal
random matrices and for useful discussions. This project was supported by the Deutsche Forschungsgemeinschaft (SFB/TR 12), 
the Nanosystems Initiative Munich (NIM) and the Center for NanoScience (CeNS). E.C. thanks the FEDER and the Spanish DGI for financial support
through Project No. FIS2007-62238.

\appendix


\section{Simplification of the Two Matrix Part of the Action}\label{ActionSimplification}

The supermatrices \eref{SupMatrixDef} written in the parametrization \eref{parametrization} can be diagonalized in each sector 
by the following transformation:
\begin{equation}
		D = \sigma^{-1} U^{\dagger} Q U \sigma^{-1} \, , \quad 
		Q = U\sigma D \sigma U^{\dagger}.
\end{equation}
Here $ \, U \, $ and $ \, U^{\dagger} \, $ are unitary block diagonal matrices:
\begin{equation}
U \equiv
		\left( \begin{array}{cccc} 
			U_R e^{i\phi^{R}} & 0 & 0 & 0 \\
			0 & U^*_R e^{-i\phi^{R}} & 0 & 0 \\
			0 & 0 & U_A e^{i\phi^{A}} & 0 \\
			0 & 0 & 0& U_A^* e^{-i\phi^{A}} \\ 
		\end{array} \right)_{RA}, 
\end{equation}
\begin{equation}
\fl U_{R/A}\equiv U(\eta^{R/A})\equiv
		\left( \begin{array}{cc} 
			\left(1-\frac{1}{2}\left(\eta^{R/A}\right)^*\eta^{R/A}\right) & -\left(\eta^{R/A}\right)^*  \\
			\eta^{R/A} & \left(1+\frac{1}{2}\left(\eta^{R/A}\right)^*\eta^{R/A}\right) \\
		\end{array} \right) \,	 ;
\end{equation}
$ \, \sigma \, $ is a symmetric orthogonal matrix
\begin{equation}
\sigma\equiv
		\left( \begin{array}{cc} 
			\sigma' & 0  \\
			0 & \sigma' \\
		\end{array} \right),
\quad
\sigma' \equiv
		\left( \begin{array}{cccc} 
			\frac{1}{\sqrt{2}} & 0 & \frac{1}{\sqrt{2}} & 0 \\
			0 & 1 & 0 & 0 \\
			\frac{1}{\sqrt{2}} & 0 & -\frac{1}{\sqrt{2}} & 0 \\
			0 & 0 & 0& 1 \\ 
		\end{array} \right) \, ;\label{sigmamatrix}	
\end{equation}
and $ \, D \, $ is a matrix which is diagonal in each sector
\begin{equation}
D\equiv
		\left( \begin{array}{cc} 
			\lambda_{R}\lambda_{R} & - \lambda_{R}\lambda_{A}  \\
			\lambda_{A}\lambda_{R} & -\lambda_{A}\lambda_{A} \\
		\end{array} \right) \otimes
		\left( \begin{array}{cccc} 
			1 & 0 & 0 & 0 \\
			0 & 0 & 0 & 0 \\
			0 & 0 & 0 & 0 \\
			0 & 0 & 0& 0 \\ 
		\end{array} \right).\label{diagmatrix}
\end{equation}
In this representation $R$ labels the retarded and $A$ the advanced sectors of the supermatrix. Each of the $4\times 4$ submatrices of the sectors in Eqs. \eref{sigmamatrix},\eref{diagmatrix}  contains bosonic and 
fermionic sectors in a pattern determined by the outer product of the super vectors defined in Eq. \eref{svectors}.
The supertrace in the two matrix part of the action can be simplified by using the invariance of the supertrace under cyclic permutations
\begin{equation}
\fl	\textrm{Str}\left[Q_iQ_j\right] =
						\textrm{Str}\left[ U_i\sigma D_i \sigma U^{\dagger}_i U_j\sigma D_j \sigma U^{\dagger}_j \right] =
								\textrm{Str}\left[ \sigma U^{\dagger}_j U_i\sigma D_i \sigma U^{\dagger}_iU_j\sigma D_j  \right],
\end{equation}
and the following property of $2\times 2$ supermatrices (cf. \cite{Zirnbauer-1986})
\begin{equation}
	U^{\dagger} \left( \eta_i \right) U \left( \eta_j \right)
										=
											U \left( \eta_j-\eta_i \right) e^{ \frac{1}{2} \left( \eta_i^* \eta_j - \eta_j^* \eta_i \right) }.
\end{equation}
This allows one to obtain Eq. \eref{Simpl_Two_Matrix_action}.


\section*{References}
\bibliographystyle{unsrt.bst}
\bibliography{PublicationCitations}
\end{document}